# Plasmon resonance in a sub-THz graphene-based detector: theory and experiment

I.M. Moiseenko (https://orcid.org/0000-0001-8933-2834)[a,*], E. Titova[a], M. Kashchenko[a], D. Svintsov[a,**]

[a] *Laboratory of 2d Materials for Optoelectronics, Moscow Institute of Physics and Technology, Dolgoprudny 141701, Russia,* *e-mail: moiseenko.im@mipt.ru*

**e-mail: svintcov.da@mipt.ru*

**Abstract**— We present a combined experimental and theoretical study of photovoltage generation in a bilayer graphene (BLG) transistor structure exposed to subterahertz radiation. The device features a global bottom and split top gate, enabling independent control of the band gap and Fermi level, thereby enabling the formation of a tunable p-n junction in graphene. Measurements show that the photovoltage arises primarily through a thermoelectric mechanism driven by heating of the p-n junction in the middle of the channel. We also provide a theoretical justification for the excitation of two-dimensional plasmons at a record-low frequency of 0.13 THz, which manifests itself as characteristic oscillations in the measured photovoltage. These plasmonic resonances, activated by a decrease in charge carrier concentration due to opening of the band gap, lead to a local enhancement of the electromagnetic field and an increase in the carrier temperature in the junction region. The record-low frequency of plasmon resonance is enabled by the low carrier density achievable in the bilayer graphene upon electrical induction of the band gap.

## INTRODUCTION

Terahertz (THz) radiation (0.1-10 THz) is a promising tool for numerous applications, including 6G wireless communications, non-invasive diagnostics, and high-resolution spectroscopy [1, 2]. The development of THz technologies is hindered by the lack of highly sensitive, low-noise, and compact detectors. Graphene and its few-layer modifications, due to unique characteristics, are one of the most promising materials for creating such detectors, in part due to their plasmonic properties [3,4]. However, the absence of a bandgap in a monolayer graphene has traditionally limited its radiation response and resulted in low efficiency of bolometric and thermoelectric rectification effects, which are critical for the operation of thermal detectors. Furthermore, the lack of established chemical doping technologies for two-dimensional materials [5] stimulates the search for new detector architectures operating at zero bias. Experimental confirmation of the photovoltage enhancement upon bandgap opening was recently demonstrated in bilayer graphene (BLG) - a unique material with an electrically tunable bandgap. It was shown that at cryogenic temperatures, the sub-terahertz sensitivity of p-n junctions in BLG significantly increases when the bandgap is increased to tens of meV [6]. Another

`

interesting result in recent years is the use of plasmonic effects in graphene for THz detection [7], even at room temperature [8].

In this work, we develop a model for sub-THz photo-thermoelectric effect in bilayer graphene transistors with gate-induced p-n junction. The model includes the plasmonic effects, i.e. possible enhancement of local sub-THz fields at particular frequencies and carrier densities. The model is compared with recent experimental data on sub-THz detection in such structures [7] and captures the observed plasmon resonance. The experimentally studied structure contains a split top gate over graphene and a global back gate, allowing for independent tuning of the bandgap and Fermi level in graphene, as well as the formation of a p-n junction within it (Fig. 1). Effective creation of a bandgap in bilayer graphene is achieved by using hafnium dioxide $HfO_2$ as the gate dielectric. A thin layer of $HfO_2$ was deposited by atomic layer deposition on a silicon substrate, which served as the back gate being transparent at sub-THz frequencies. Next, the hBN-graphene-hBN layered structure was transferred onto the $HfO_2$ surface, after which Ti/Au contacts and split gates were deposited. Hereafter, we will refer to the top gates as right (x>0) and left (x<0), and the graphene regions under the corresponding gates as left and right, respectively.

To focus terahertz radiation on the micrometer-size graphene-based transistor, we used a bow-tie antenna connected to the graphene channel. The radiation source was an impact ionization avalanche transit-time (IMPATT) diode operating at a nominal frequency of 0.13 THz with an output power of 16.4 mW. Considering transmission losses through the focusing lens, cryostat windows, and attenuator system, the power reaching the antenna area of the detector was estimated to be $P_{THz} \approx 70$nW. The electrical formation of a band gap in graphene was confirmed by an exponential increase in graphene resistance at the charge neutrality point as the average perpendicular electric field increased, the latter controlled by the top and back gate voltages. The graphene channel resistance was measured in a two-terminal configuration using a low-amplitude alternating current ($I_{sd} \approx 25$ nA, frequency 83 Hz) applied between the source and drain electrodes. All photovoltaic measurements were carried out in zero-bias mode at a cryogenic temperature of T=7 K using a standard lock-in detection technique with a modulation frequency of about 14.5 Hz

THEORY OF PHOTOVOLTAGE GENERATION BY THERMOELECTRIC EFFECT

In our theoretical model, we assume that the photovoltage in graphene arises from a thermoelectric effect caused by heating of charge carriers along the transistor channel. This is evidenced by the experimentally observed sixfold sign reversal of the photovoltage with the gate sweep, which is a distinctive feature of the thermoelectric mechanism of photovoltage generation [7]. As will be shown later, the maximum heating of charge carriers in graphene occurs near the p-n junction, created at the boundary between the right and left parts of graphene by applying gate voltages $V_g$ of different signs to the right and left top gates. The sign of the gate voltage determines the type of conductivity in graphene under that gate (electron conductivity for $V_g>0$ and hole conductivity for $V_g<0$). To estimate the thermoelectric photovoltage between the source-drain contacts, we use the expression

$$V_{ph} = (S_L - S_R)\Delta T , \quad (1)$$

where $S_L$ and $S_R$ are the Seebeck coefficients in the right and left regions of graphene, and $\Delta T$ is the radiation-induced heating of the p-n junction formed in the channel. The Seebeck coefficient is determined by the kinetic coefficients for electrons and holes $\alpha_{e/h}$ and $\sigma_{e/h}$ as

$$S = (\alpha_e - \alpha_h)/(\sigma_e + \sigma_h), \qquad (2)$$

where $\alpha$ relates the current density in graphene to the temperature gradient $J = \alpha \nabla T$, and $\sigma$ is the electrical conductivity, defined as

$$\alpha_e(E_F) = \frac{e}{2k}\int_{E_C}^{\infty} \rho(E)v^2(E)\tau \frac{E-E_F}{kT}\frac{\partial f_0}{\partial E}dE,$$

$$\sigma_e(E_F) = -\frac{e^2}{2}\int_{E_C}^{\infty} \rho(E)v^2(E)\tau \frac{\partial f_0}{\partial E}dE,$$

where $\rho(E)$ is the density of states, $\tau$ is the momentum relaxation time, $v$ is the modulus of band velocity, and $f_0$ is the equilibrium Fermi function. The transport coefficients for holes can be found from $\alpha_h(E_F) = \alpha_e(-E_F)$, $\sigma_h(E_F) = \sigma_e(-E_F)$. For further calculations, we assume $\rho(E) = \text{const}$, $\tau=\text{const}$, the resulting dependence of $S$ on band gap weakly depends on these assumptions. The sign of the Seebeck coefficient is directly related to the sign of the gate voltage and indicates the type of majority charge carriers drifting along graphene, corresponding to $S<0$ for electrons and $S>0$ for holes (Fig. 2(a)).

Applying voltage to the back gate results in the appearance of a band gap in graphene, which increases approximately proportionally to the back gate voltage. With increasing bandgap, the Seebeck coefficient grows at a fixed Fermi energy in the case of $E_F<E_g/2$ due to the suppression of the bipolar contribution to thermoelectric transport. and is practically independent of the bandgap in the case of $E_F>E_g$ (Fig. 2(a)). In a structure with graphene containing both electron and hole conductivity regions simultaneously, the situation is more complex, since the photovoltage magnitude is determined by the difference in Seebeck coefficients in graphene under the right and left gates (Fig. 2(b)). The symmetry and sign-changing pattern of the calculated $S_R-S_L$ matches the experimentally measured photovoltage which, again, confirms the dominant role of thermoelectric effect in the photoresponse.

Next, to determine the photovoltage, it is necessary to evaluate the heating of charge carriers in graphene by the incident sub-THz radiation. The temperature along the graphene is determined by solving the heat balance equation on $T(x)$, which we adopt in the following form:

$$\Delta T(x)C_e\gamma_e - \frac{\partial}{\partial x}\left(\chi(x)\frac{\partial \Delta T(x)}{\partial x}\right) = \frac{1}{2}\text{Re}\,\sigma_\omega(x)|E_x(x)|^2, \quad j=L, R. \qquad (3)$$

where $\chi(x)$ is the thermal conductivity of graphene, $\gamma_e = 1/\tau_e$ is the inverse cooling time due to the interaction with the substrate, $\mathrm{Re}\,\sigma_\omega(x)$ is the real part of ac conductivity evaluated at sub-THz frequency, and $E_x(x)$ is the local sub-THz electric field. The equation states the balance of ac Joule heating (right hand side) and the heat losses into the substrate (first term in the left-hand side) and into the metal contacts (second term in the left-hand side). All material constants of graphene, such as thermal and electrical conductivity, are assumed piecewise constant in the left and right sections, respectively. The respective constant values are amended with subscripts L and R for the left and right sections. Solutions to equation (3) for the temperature in the right and left parts of graphene were obtained using boundary conditions for local equality of temperatures $\Delta T_L(0) = \Delta T_R(0)$ and heat fluxes $\chi_L \partial \Delta T_L(x)/\partial x|_{x=0} = \chi_R \partial \Delta T_R(x)/\partial x|_{x=0}$ on the p-n junction. It appears possible to find a semi-analytical solution of (3) in terms of the Green's function G(x,x₀) of the heat balance equation. The right-hand side of equation (3) contains the electric field E(x) induced in the structure by an external electromagnetic wave. This field is largely different from the incident one due to the self-consistent (plasmonic) effects. To calculate the distribution of the oscillating field, we use the continuity equation in the form

$$-i\omega\rho(x) + \frac{\partial J(x)}{\partial x} = 0, \tag{4}$$

where the surface charge density in the local capacitance approximation is $\rho(x) = C\varphi(x)$, where $C = \varepsilon\varepsilon_0/d$ is the areal capacitance between graphene and a gate located at a distance d, and the surface current density in graphene is related to the electric field through Ohm's law $J(x) = \sigma_\omega(x)E(x)$, where $E(x) = -\partial\varphi(x)/\partial x$ is the sought-for electric field, and the conductivity of graphene is, again, assumed piecewise constant $\sigma_\omega(x) = \sigma_L\theta(-x) + \sigma_R\theta(x)$. Equation (4) was solved with boundary conditions for potentials continuity $\varphi_L(0) = \varphi_R(0)$ and currents $\sigma_L \partial \varphi_L(x)/\partial x|_{x=0} = \sigma_R \partial \varphi_R(x)/\partial x|_{x=0}$ at the p-n junction. The potentials at the source and drain contacts were considered fixed by the antenna, $\varphi(\pm L/2) = \pm V_{ant}/2$. The solution of equation (4) with respect to the oscillating potential has the form

$$\varphi_j(x) = \frac{V_{ant}}{2}\csc\left[L(q_L+q_R)/4\right]\sin\left[L(q_L-q_R)/4 + q_j x\right], \tag{5}$$

where $q_j = \left(\sqrt{\frac{\sigma_j d}{i\omega\varepsilon\varepsilon_0}}\right)^{-1}$ the wave vector of plasmons in the j-th section (j={L, R}) of the graphene channel. We estimated the voltage developed between antenna arms based on the power of the source $P_{THz}$ and the impedance of the antenna $Z_{ant}$ as $V_{ant} = \sqrt{Z_{ant}P_{THz}}$. The charge carrier temperature profile in graphene is symmetrical with respect to the p-n junction in the case of equal electron and hole concentrations in the right and left sections and has a maximum at the center (i.e. at the p-n junction). Its magnitude reaches about a few fractions of a Kelvin for given THz power of several nanowatts (Fig. 3a). Thermal diffusion occurs from the heated center toward the cold contacts of the source and drain, the respective heat diffusion length is $L_{T,j} = \sqrt{\chi_j / \gamma_e C_e}$, $j = L, R$. Changing the Fermi energy under the gates alters the carrier diffusion length, which shifts the temperature maximum position under one of the gates.

One of the most remarkable results of the developed model is the prediction of the oscillatory structure in the dependence of junction temperature on the parameters of graphene – its band gap and carrier density (Fig. 3(b)). These peaks are associated with the excitation of two-dimensional plasmons under the gates. This effect leads to a local increase in the electric field, to increase in local Joule heating density, and eventually – to an increase in the temperature at the p-n junction. Given the relatively large channel length (L=6 μm), the excitation of two-dimensional plasmons becomes possible due to a decrease in the charge carrier concentration in graphene with an increase in the band gap. The frequency of two-dimensional plasmons $\omega_{pl} \propto \sqrt{n_0(\varepsilon_F, \varepsilon_g)}$ can be shifted on sub-THz frequencies at the charge carriers concentration in graphene $n_0(\varepsilon_F, \varepsilon_g) \approx 10^{11}$ cm$^{-2}$, and the wavelength of plasmons becomes comparable to the large length of the detector channel even at sub-THz frequencies.

Finally, the obtained expressions for the Seebeck coefficients under the right and left gates (2), as well as the obtained temperature distribution along the graphene channel, allow us to estimate the thermoelectric photovoltage in graphene using equation (1). We investigated the photovoltage for a fixed Fermi energy under the left gate as a function of the magnitude and sign of the Fermi energy under the right gate for various band gap values in graphene (Fig. 4a). We examined the Fermi energy region near the neutrality point of graphene, where the Seebeck coefficient reaches its maximum value. The magnitude of the photovoltage increases with increasing band gap and reaches its maximum value at different signs of the Fermi energy in the right and left parts of graphene, and the sign of the photovoltage is determined by the sign of the difference in the Seebeck coefficients $(S_R - S_L)$. As the Fermi energy approaches the edge of the energy band, characteristic plasmonic oscillations in the photovoltage become increasingly pronounced near its maximum values. This occurs for both electron ($E_{FR}>0$) and hole ($E_{FR}<0$) conductivity under the right gate (Fig. 4). The calculated photovoltage agrees well with experimental results from photovoltage measurements in a bilayer graphene-based structure (Fig. 4b).

To conclude, we have provided a theoretical model for the first experimental observation of graphene plasmons [7] at a very low frequency of 0.13 THz. Previously, it was believed that the main obstacle to observing plasmon resonance in the technically desirable sub-THz frequency range is the strong damping of plasmons at these frequencies ($\omega\tau < 1$). In this study, we show that the presence of residual carriers in graphene at charge neutrality (activated thermally or due to local electric potential fluctuations) is not less serious obstacle. These residual carriers pin the resonant frequency at quite large values. As we have shown, low carrier densities and low resonant frequencies can be achieved in bilayer graphene with electrically induced band gap.

The presented theoretical model can be improved in several aspects to achieve better agreement with experimental data. First of all, realistic device geometry with a relatively wide slit between gates can be taken into account. Second, the momentum and energy relaxation times can be evaluated microscopically assuming electron scattering by impurities and energy loss via electron-optical phonon interactions. Finally, rectification by metal-graphene Schottky junctions can be considered in line with rectification by the central p-n junction. Despite these approximations, the developed simple model captured the main experimental finding of [7], the manifestation of plasmonic oscillation in photovoltage at large induced band gap.

## CONCLUSIONS

We have shown that plasmonic effects play an important role in the sub-terahertz rectification by p-n junctions in bilayer graphene even at very low sub-terahertz frequencies (f=130 GHz). The developed model captures the experimentally observed plasmonic oscillation of photovoltage [7] emerging at large induced bandgap in bilayer graphene. Induction of the gap reduces the density of carriers and lowers the resonant frequency to hundreds of GHz. The results of the study open up prospects for the creation of highly sensitive, compact, and tunable terahertz detectors based on bilayer graphene and other two-dimensional materials with adjustable band gaps. Further development of theory and experiments in this direction will allow us to optimize the geometry and electrical parameters of devices, as well as bring their widespread practical application closer.

## FUNDING

The work was financially supported by the grant №21-79-20225 - P of the Russian Science Foundation.

## CONFLICT OF INTEREST

The authors declare that they have no conflicts of interest.

FIGURE CAPTIONS

**Fig. 1.** Schematic view of the studied structure.

**Fig. 2** (a) Seebeck coefficient as a function of Fermi energy for different bandgaps $E_g$=0, 50 meV, and 80 meV and (b) Difference in Seebeck coefficients between the right and left graphene regions as a function of the Fermi energies in these regions.

**Fig. 3** (a) Spatial profile of the charge carrier temperature along the graphene with a p-n junction in the center of the channel of length L (at x=0) for $E_{FL}$=50 meV and different values of $E_{FR}$ with a width of $E_g$=100 meV. (b) Dependence of the temperature in the p-n junction on the band gap at a fixed value of the Fermi energy under the left gate $E_{FL}$=50 meV and different values of $E_{FR}$ under the right gate.

**Fig. 4** (a) Photovoltage calculated as a function of the Fermi energy in graphene under the right gate at a fixed Fermi energy under the left gate $E_{FL}$=50 meV for different values of the band gap in graphene. (b) Photovoltage measured as a function of the voltage on the right gate for different values of the voltage on the lower gate corresponding different bandgaps.

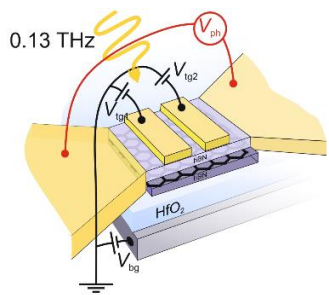

Fig. 1

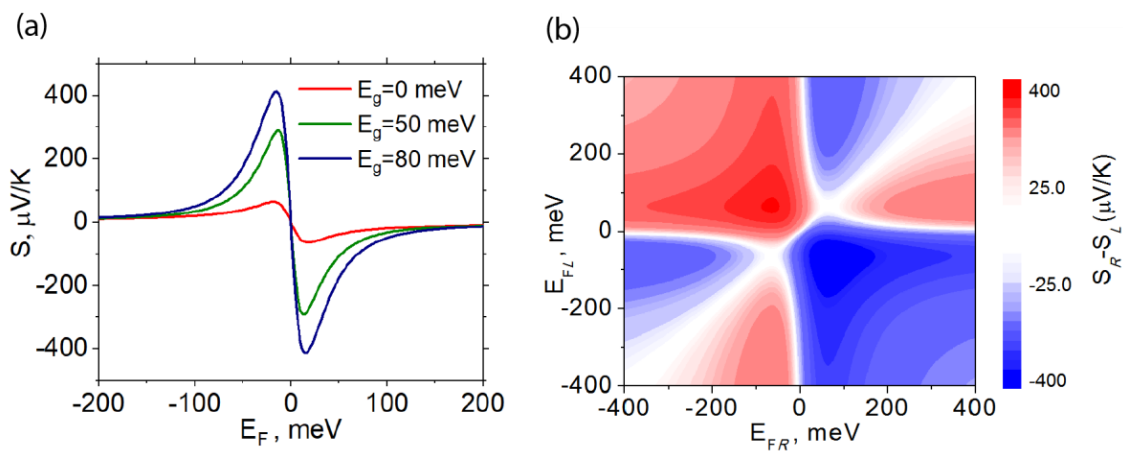

Fig. 2.

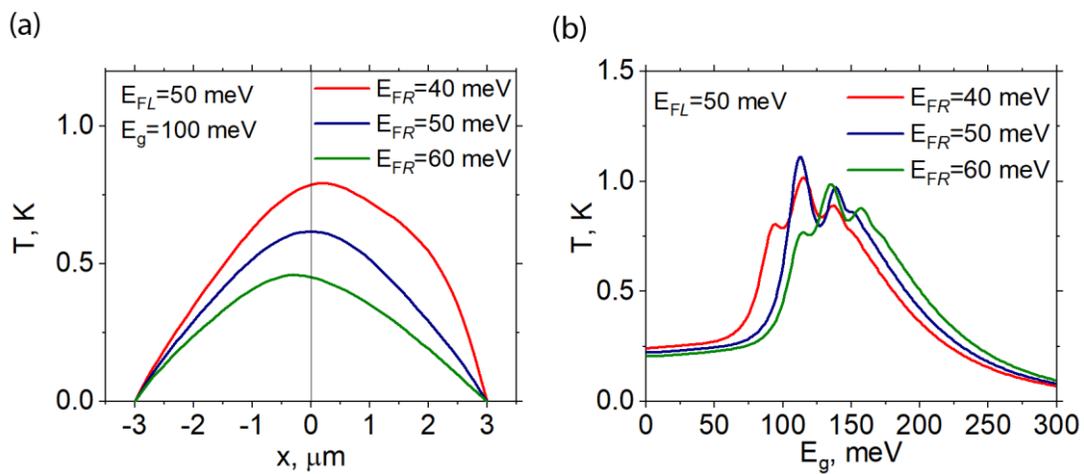

Fig. 3.

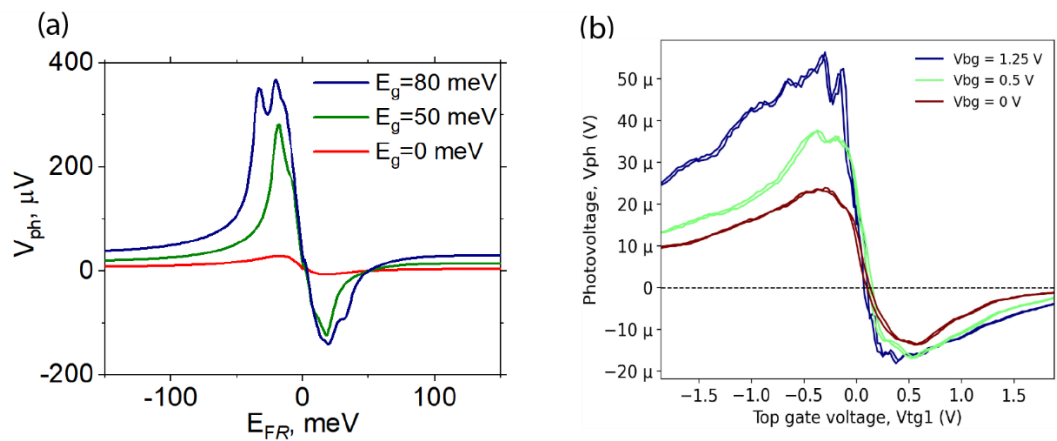

Fig. 4.